\documentclass[useAMS,usenatbib]{mn2e}
\usepackage{graphicx,amssymb,epsfig}
\usepackage{aas_macros}
\usepackage{natbib}

\newcommand{\msun}{\,{\rm M_\odot}}

\newcommand{\beq}{\begin{equation}}
\newcommand{\eeq}{\end{equation}}
\newcommand{\ba}{\begin{eqnarray}}
\newcommand{\ea}{\end{eqnarray}}
\def\spose#1{\hbox to 0pt{#1\hss}}
\newcommand{\lta}{\mathrel{\spose{\lower 3pt\hbox{$\mathchar"218$}}
      \raise 2.0pt\hbox{$\mathchar"13C$}}}
\newcommand{\gta}{\mathrel{\spose{\lower 3pt\hbox{$\mathchar"218$}}
      \raise 2.0pt\hbox{$\mathchar"13E$}}}
\def\simlt{\mathrel{\rlap{\lower 3pt\hbox{$\sim$}}\raise 2.0pt\hbox{$<$}}}
\def\simgt{\mathrel{\rlap{\lower 3pt\hbox{$\sim$}} \raise 2.0pt\hbox{$>$}}}

\title[Compact massive objects in Virgo galaxies: the black hole population]
{Compact massive objects in Virgo galaxies: the black hole population}

\author[Volonteri, Haardt \& G\"{u}ltekin]{Marta Volonteri$^{1}$, Francesco Haardt$^{2}$
\& Kayhan G\"{u}ltekin$^{1}$\\
\footnotemark[1] 
$^{1}$ Department of Astronomy, University of Michigan, Ann Arbor, MI, USA\\
$^{2}$ Dipartimento di Fisica e Matematica, Universit\'a dell'Insubria, Via Valleggio 11, 22100 Como, Italy}
\begin{document}
\maketitle
\begin{abstract}
We investigate the distribution of massive black holes (MBHs) in the Virgo cluster. Observations suggest that AGN activity is widespread in massive galaxies ($M_*\simgt 10^{10}\msun$), while at lower galaxy masses star clusters are more abundant, which might imply a limited presence of central black holes in these galaxy-mass regimes. We explore if this possible threshold in  MBH hosting, is linked to {\it nature}, {\it nurture}, or a mixture of both. The {\it nature} scenario arises naturally in hierarchical cosmologies, as  MBH formation mechanisms typically are efficient in biased systems, which would later evolve into massive galaxies. {\it Nurture}, in the guise of  MBH ejections following  MBH mergers, provides an additional mechanism that is more effective for low mass, satellite galaxies. The combination of inefficient formation, and lower retention of  MBHs, leads to the natural explanation of the distribution of compact massive objects in Virgo galaxies.  If MBHs arrive to the correlation with the host mass and velocity dispersion during merger-triggered accretion episodes, sustained tidal stripping of the host galaxies creates a population of MBHs which lie above the expected scaling between the holes and their host mass, suggesting a possible environmental dependence.  
\end{abstract}

\section{Introduction}
The nearby Virgo cluster is a perfect laboratory to investigate the evolution of galaxies in a dense environment. Recently, observations of a large sample of galaxies suggested that the properties of nuclei, either quiescent or active, in Virgo galaxies are strongly mass-dependent. This latter finding is in very good agreement with the general trend found also in the SDSS, where very few AGN are found in galaxies with stellar mass $M_*< 10^{10}\msun$\citep{Kauffmann2003, Kewley2006}.

Decarli et al. (2007) analyzed nuclear activity in late type galaxies in the Virgo cluster. They conclude, quite remarkably, that at galaxy mass\footnote{Decarli et al. (2007) measure the dynamical mass of the galaxy within the optical radius, determined at the $25^{th}$ mag arcsec$^{-2}$ isophote in the B band. This is an upper limit to the stellar mass of galaxies, but a lower limit to the total baryonic mass.} $M_{gal}> 10^{10.5}\msun$ the AGN fraction is unity. As a central black hole is a necessary condition for AGN activity, we conclude that the black hole occupation fraction (BHOF)  must be unity as well. \cite{Cote2006}, Wehner \& Harris (2006) and Ferrarese et al. (2006) find that, below $M_*\sim 10^{10}\msun$, Virgo galaxies exhibit nuclear star clusters, whose mass scales with $M_*$  in the same fashion as those of the massive black holes detected in brighter galaxies \citep{Magorrian1998, MarconiHunt2003, Haring2004}. Although the existence of a nuclear star cluster does not rule out a small, hidden  MBH, it is suggestive that Ferrarese et al. (2006) conclude that ``bright galaxies often, and perhaps always, contain supermassive black holes but not stellar nuclei. As one moves to fainter galaxies, nuclei become the dominant feature while  MBHs might become less common and perhaps disappear entirely at the faint end."  

There are three interlaced
sides of the intriguing story which appears to link stellar nuclei and
MBHs: 1) understand if (and why) MBHs populate preferentially bright
galaxies; 2) understand why stellar nuclei populate preferentially
faint galaxies and 3) understand why the ratio of nuclear to galaxy
mass is identical to the ratio of MBH to galaxy mass.  In this paper we address the first issue. A possible hint to explain the predominance of star clusters in small galaxies may come from comparing the dynamical time scale and the fragmentation time scale of the infalling gas. Detailed calculations are necessary to test this hypothesis, but it can be argued that in shallow potentials gas could fragment before reaching the center of the galaxy. This is even more suggestive in the case of merger-induced gaseous infall. Regarding the third issue, \cite{Emsellem2007} for instance show that if galaxies have Sersic profiles, radial compressive forces trigger the collapse of gas in the central regions, and the mass of the nuclear cluster that forms is about 0.1\% of the mass of the host galaxy. So, nuclear cluster formation and MBH feedback might produce the same scaling relation with galaxy mass, but it is unclear whether this is a coincidence or the result of a single, unexplored process.

Volonteri et al. (2007) investigate the overall distribution of  MBHs in galaxies, as a function of the host velocity dispersion, $\sigma_*$ , as traced by the host halo (cfr. Ferrarese 2002), and find that the efficiency of  MBH formation decreases with halo mass, and isolated dwarf galaxies are most likely to be devoid of a central  MBH. This is a common feature of  MBH formation models which invoke gas-dynamical processes in the high redshift Universe, either via direct collapse \citep[e.g.,]{haehnelt1993,LoebRasio1994,Eisenstein1995,BrommLoeb2003,BegelmanVolonteriRees2006,LN2006}, or via an intermediate stage of metal free Population III star \citep[e.g.,]{CBA84,MadauRees2001,VHM}. The common feature is the need for deep potential wells at cosmic epochs when the protogalaxy population was dominated by minihalos ($M_h<10^7\msun$). \\
\indent A second physical phenomenon strengthens the likelihood that dwarf galaxies, even if seeded with a  MBH at early times, are now lacking one: the gravitational recoil. That is, the general-relativistic effect which imparts velocity to the center of mass of a merging  MBH binary, when the  MBHs plunge into each other's last stable orbit by emission of gravitational radiation. Gravitational waves carry, in general,  a non-zero net linear momentum, which establishes a preferential direction for the propagation of the waves. As a consequence, the center of mass of the binary recoils in the opposite direction \citep{redmountrees}, possibly causing the ejection of MBHs from the potential wells of their host galaxies  \citep[e.g.,][]{Madau2004,MadauQuataert2004, Merrittetal2004, VolonteriRees2006, Haiman2004, SchnittmanBuonanno2007}. The progress of numerical relativity is now leading to a convergence in the estimates of the recoil. 
Schwarzschild, i.e., non-spinning, black holes \citep[e.g.,][]{Bakeretal2006} are expected to recoil with velocities below 200 $\rm{km\,s^{-1}}$, and a similar range is expected for black holes with low spins, or with spins (anti-)aligned with the orbital axis. However, when the spin vectors have opposite directions and are in the orbital plane, the recoil velocity can be as large as a few thousands $\rm{km\,s^{-1}}$ \citep{Campanelli2007b,Gonzalez2007,Campanelli2007}.

Schnittman (2007) shows that the hierarchical nature of galaxy assembly implies that ejections do not lead to void nuclei, even in the high-recoil limit. This is weaved into the very nature of galaxy assembly, via a series of mergers, in a pattern typically dubbed ``merger tree". The botanical essence of the tree implies that many branches converge into a central trunk, so that in every generation of the tree, the number of galaxies decreases, while the fraction of  MBHs can increase, even if ejections operate. This is especially true for  ``central galaxies", which represents the trunk of the merger tree. 

The situation is different for satellites in a galaxy cluster: they correspond to loose branches that do not merge into the main trunk. Satellites are indeed galaxies that enter the dark matter halo of the cluster, but do not merge with the central galaxy. 

In this paper we develop simple models that describe the dynamical evolution of satellite galaxies in a cluster, focusing on the fate of their nuclei. We estimate the influence of  MBH formation mechanisms, and of  MBH ejections for shaping the BHOF in the Virgo cluster today.

\section{MBH formation and dynamical evolution}
We follow the evolution of the  MBH population in a $\Lambda$CDM Universe along the history of a cluster-size halo. Our technique and cosmological framework is similar to the one described in \cite{VHM}. However we now focus on the merger history of halos which have a total mass $M_h=10^{15}\msun$ at $z=0$ \citep[see also][]{Yoo2007}.  We track the dynamical evolution of  MBHs ab-initio and follow their assembly down to $z = 0$. We adopt very simplified assumptions on  MBH formation and mass growth, based on a phenomenological approach. Our goal is to highlight the influence of two main processes, efficiency of  MBH formation and dynamics, rather than develop an omni-comprehensive evolutionary model plagued by free parameters. 

Several theoretical arguments indicate that  MBH formation proceeds at very high redshift, and probably in biased halos \citep[e.g.,][]{MadauRees2001,Haiman2004,Shapiro2005,VolonteriRees2005,VolonteriRees2006}. Additional parameters, such as the angular momentum of the gas, its ability to cool and its metal enrichment, are likely to set the exact efficiency of  MBH formation and the redshift range when the mechanism operates \citep[for a thorough discussion see][]{Volonterietal2007}. 

We explore here 4 simplified models: a very high efficiency model (model I, ``higheff"), where all halos with a formation redshift $z>5$ host a  MBH, an intermediate efficiency model (model II, ``mideff"), where all halos with a formation redshift $z>12$ host a  MBH, and finally two biased, low efficiency model, where MBHs form only in halos  which represent  density peaks with $\nu_c=3$\footnote{$\nu_c=\delta_{\rm c}(z)/\sigma(M,z)$ is the number of standard deviations which the critical collapse overdensity represents on mass scale $M$. $\sigma(M_h,z)$ is the root mean square fluctuation of the linear density field at redshift $z$, and $\delta_c(z)$ is the threshold density for collapse of a homogeneous spherical perturbation at redshift z. See, e.g., Peebles 1993}(model III, ``peak3") or $\nu_c=3.5$ (model IV, ``peak3.5") and have a formation redshift $z>12$. The choice $\nu_c=3.5$ in model IV is based on the results in \cite{VHM}, and is meant as a hard threshold to illustrate the effect of biased formation.  Our models link the formation of a MBH seed to the halo mass and formation redshift of a given galaxy only, and we do not assume here that MBHs are {\it formed} in bulges. Several models of MBH formation \citep[e.g.,][]{ Eisenstein1995, Koushiappas2004, BegelmanVolonteriRees2006, LN2006} additionally require that MBH seed hosts have low angular momentum, thus hindering the existence of MBHs in pure disc galaxies.

Our four ``models" of  MBH formation define how often and when halos are populated with MBHs. They provide the initial occupation fraction, and they set the minimum mass of MBH hosts today. The higher the redshift when  MBH formation ceases, the larger the typical  MBH host today is. This is a direct consequence of hierarchical cold dark matter models \citep[cfr.][]{MadauRees2001}.

Additionally we have to follow the dynamical evolution of halos and embedded  MBHs all the way to $z=0$ in order to determine the occupation fraction of  MBHs and their properties. We focus here on the effect that  MBH ejections, namely due to gravitational waves recoils, have on the properties of the  MBH population at $z=0$. The magnitude of the recoil depends on the mass ratio of the merging  MBHs, the spins of the  MBHs, the orbital parameters of the binary. 

First, to evaluate mass ratios, we have to model the mass-growth of  MBHs. We base our modeling on plausible assumptions, supported by both simulations of AGN triggering and feedback \citep{Springel2005b}, and analysis of the relationship between  MBH masses ($M_{BH}$) and the properties of their hosts \citep{Mclure2004,Wyithe2005}. \cite{Wyithe2005} show that if the relationship between the mass of a  MBH and the velocity dispersion of the host found for local galaxies \citep{Tremaine2002,Ferrarese2000,Gebhardt2000} does not evolve with redshift, then the correlation between the masses of  MBHs and their hosts evolves with redshift in a way compatible with observational results by \cite{Mclure2004}. Additionally, \cite{Springel2005b} suggest that the ditto $M_{BH}-\sigma_*$ relation is established during galaxy mergers that also fuel  MBH accretion and form bulges. We therefore assume that after every merger between two galaxies with a mass ratio larger than $1:10$, their MBHs attain the mass predicted by the $M_{BH}-\sigma_*$ for each of the merging galaxies. 

Hence, although  in our models the presence of a MBH is not uniquely coupled with bulge formation, the mass of a black hole is set during the same event that forms the  host bulge. Accordingly, when a binary of  MBHs merge, their mass ratio scales with the $M_{BH}-\sigma_*$ relation appropriate for the velocity dispersions of the progenitor halos of the MBHs. Note that if the $M_{BH}-\sigma_*$ relation scales with redshift as suggested by, e.g. \cite{Woo2006, Treu2004} the mass ratio of merging binaries would be unchanged, so the occupation fraction of black holes would not be affected. We further assume that  MBHs merge within the merger timescale of their host halos, which is a likely assumption for  MBH binaries formed after gas rich galaxy mergers \citep{Escalaetal2004,Dottietal2006,Dotti2006c}.

To determine the efficiency of  MBH ejections due to gravitational recoils, we need also information on the magnitude and orientation of  MBH spins at the time of the merger. We will express  MBH spins as a function of the dimensionless parameter $\hat a \equiv J_h/J_{max}=c \, J_h/G \, M_{\rm  MBH}^2$, where $J_h$ is the angular momentum of the black hole. 

Non-spinning  MBHs, or binaries where  MBH spins are aligned with the orbital angular momentum are expected to recoil with velocities below 200 $\rm{km\,s^{-1}}$. The recoil velocity is largest for  MBHs with large spins, when the spin vectors have opposite directions and are in the orbital plane \citep{Campanelli2007b,Gonzalez2007,Campanelli2007}. Assuming that BHs at the time of the merger always have antialigned spins in the orbital plane \citep[as in][]{Volonteri2007} would provide a strict upper limit to the effect of the recoil. 
However, the configuration yielding the highest recoil velocities is probably rather uncommon, as pointed out by \cite{Bogdanovic2007}. \cite{Bogdanovic2007} suggest that when the  MBH merger happens in a gas rich environment, and is accompanied by accretion, the most likely configuration has spins aligned (or anti-aligned) with the orbital angular momentum, thus avoiding the highest recoil velocity. Conversely, in gas poor mergers there is no preferential spin alignment, so all spin/orbital parameters configurations are equally probable.  We will assume in the following that orbital parameters and spin configuration are isotropically distributed, likely providing a soft {\it upper limit} to the strength of the recoil. 

The distribution and most probable value of  MBH spins is observationally largely unconstrained. There is evidence that  MBHs in some local AGN galaxies do spin \citep{Streblyanska2005,Comastri2006,Brenneman2006}, based on iron line profiles \citep{Miller2007,Fabian1989,Laor1991}. High spins in bright quasars are also indicated by the high radiative efficiency of quasars, as deduced from observations, by applying Soltan's argument \citep [and references therein ]{Soltan1982,Wang2006}. However, neither observation firmly establishes that most  MBHs have large spins, although there are theoretical arguments to expect so \citep{Moderski1998,Volonterietal2005} as spin-up is a natural consequence of prolonged disc-mode accretion for any hole that has (for instance) doubled its mass by capturing material with constant angular momentum axis \citep{Bardeen1970,Thorne1974}. \cite{King2005} argue instead that most  MBHs have very low or no spin, due to preferential accretion of counter-rotating material, or to short-lived accretion episodes \citep{King2005}. Waiting for additional observations\footnote{The spin is a measurable parameter, with a very high accuracy, in the gravitational waves {\it LISA} signal 
\citep{BarackCutler2004,Berti2005,Lang2006,Vecchio2004}. Gravitational waves emitted when a stellar mass MBH, or even a white dwarf, or neutron star, falls into a massive black hole can be used to map the spacetime of MBHs, and as a consequence,  MBH spins.}, we consider here two extreme cases, that likely allow us to bracket the typical configurations: either that all  MBHs have exactly null spin, or that all  MBHs have $\hat a=0.9$. The latter value is slightly lower than the canonical $\hat a=0.998$ \citep{Thorne1974}, but it is consistent with magnetohydrodynamical simulations of disc accretion \citep{Gammieetal2004}. We also run a control simulation where we set the recoil velocity to zero for all  MBH mergers. 
\begin{figure}   
  \includegraphics[width=0.9\columnwidth]{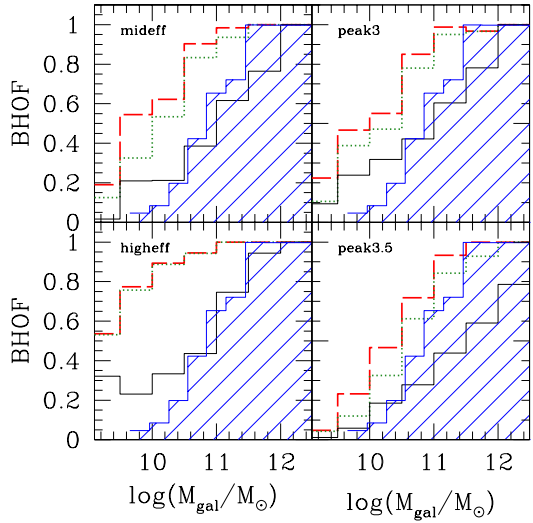}
\caption{Occupation fraction of MBHs in Virgo galaxies. Hatched histogram: Decarli et al. 2007. Dashed histogram:  no recoil. Dotted histogram:  $\hat a=0$. Solid histogram:  $\hat a=0.9$. From bottom-left panel, clockwise: model I (higheff), model II (mideff), model III (peak3), model IV (peak3.5). Model histograms are based on total baryonic masses.}
\label{fig1}
\end{figure}

For every galaxy merger we track jointly the dynamical evolution of the  MBHs and of the host halo. In addition to the dynamics of  MBH binaries, as described above, we trace the mass evolution of their hosts, including the {\it increase} due to galaxy mergers and the {\it decrease} due to mass stripping of the halo within the gravitational potential of the halo. Our treatment is very simple: we integrate the equation of motion of the satellite in the gravitational potential of the cluster (assuming a non singular isothermal sphere), including the dynamical friction term. At every step of the integration we compare the density of the satellite to the density of the cluster halo at the location of the satellite. Tidal stripping occurs at the radius within which the mean density of the satellite exceeds the density of the galaxy interior to its orbital radius \citep{Taylor2001}. We trace tidal stripping of all satellites from the time of the merger to $z=0$.

\section{Black holes occupation fraction and mass scaling}
By applying the assumptions described in Section 2 to the assembly history of galaxies in a region that can be identified as a cluster at $z=0$, we can determine the fraction of the galaxies which host a  MBH, and if there are trends with the size and mass of the host galaxy. We note here that when we refer to the mass of the host, we assume for simplicity the total baryonic mass, which we scale with the cosmic baryon fraction (14\%).  The stellar mass $M_*$ of galaxies, as well as the dynamical mass within the optical radius, $M_{gal}$, are undoubtedly non constant functions of the halo mass, environmental density, and star formation history of the satellite itself. Total baryonic masses are therefore {\it upper limits} to the measurable stellar and dynamical masses of the hosts. Given our assumptions on MBH formation and growth, the occupation fraction does not sensibly depend on bulge mass, but rather on the total mass of a galaxy, considered a tracer of the cosmological formation time (linked to the halo bias), and of the dynamical evolution (within the cluster potential). We stress, however, that pure disc galaxies are likely unsuitable for MBH formation (see discussion in section 2).

We determine the BHOF as a function of halo mass \citep[compare with the standard definition for galaxies, e.g.,][]{Peacock2000}, which is shown in Figure ~\ref{fig1} for the four different models I, II, III, IV. We compare our  BHOF to the distribution of AGN in Virgo galaxies by \cite{Decarli2007}, who find signs of AGN activity in all {\it late type} galaxies in Virgo with galaxy mass above $3\times10^{11}\msun$. Clearly the AGN occupation function is a lower limit to the  BHOF, that is at every galaxy mass:  BHOF($M_{gal}$,AGN)$\leqslant$ BHOF($M_{gal}$)$\leqslant$1.

The  BHOF of our models where the recoil velocity is set to zero informs us of the ``nature" scenario, in that only the initial conditions of  MBH formation at high redshift enter into the final  BHOF that we find at $z=0$. We see that in all models I, II, III and IV, the  BHOF is of order unity only above a certain galaxy mass. It is apparent that the existence of a threshold in galaxy mass (as traced also by the velocity dispersion) is a natural outcome of  MBH formation scenarios where the formation redshifts are pushed to high redshift. We note here that it is indeed necessary that  MBHs are formed early on, as luminous quasars are detected at redshift 5 and higher (e.g., Fan et al. 2004). Obviously, the lower the typical  MBH formation redshift, the higher the fraction of dwarf galaxies that can host a  MBH. Even without additional physical processes, our analysis shows that the possible decrease in BHOF at low galaxy mass is an expected feature in models that are based on realistic  MBH seed formation mechanisms. A similar behavior, that is, an increasing BHOF with galaxy mass, is expected also in field galaxies \citep{Volonterietal2007}, but it is exacerbated in cluster satellite galaxies, where coalescence rounds \citep[using the terminology of][]{Sesa2007} are truncated. 

The dynamical evolution of the  MBH population modifies the  BHOF in the direction of decreasing it, as  MBHs are prone to ejections, and the ejection probability depends on the galaxy mass, via the escape velocity. Although the changes are not dramatic, that is  MBH kicks do not deplete most $z=0$ galaxies of their  MBHs \citep{Volonteri2007,Schnittman2007}, the effects are non negligible, especially for galaxies with mass below $10^{11}\msun$. The  BHOF becomes a steeper function of the galaxy mass, and the threshold for BHOF close to unity shifts towards higher masses. If we compare our models to the results by \cite{Decarli2007}, who find that all galaxies with mass above $3\times10^{11}\msun$ show signs of AGN activity, hence host a central  MBH, we can rule out very high biased models (model IV, peak3.5) with high recoil velocity (solid lines in Figure ~\ref{fig1}), so if  MBH formation is indeed very biased, either  MBH spins align during the pre-merger orbital decay, as suggested by \cite{Bogdanovic2007}, or  MBHs have small spins. 

\begin{figure}   
  \includegraphics[width=0.9\columnwidth]{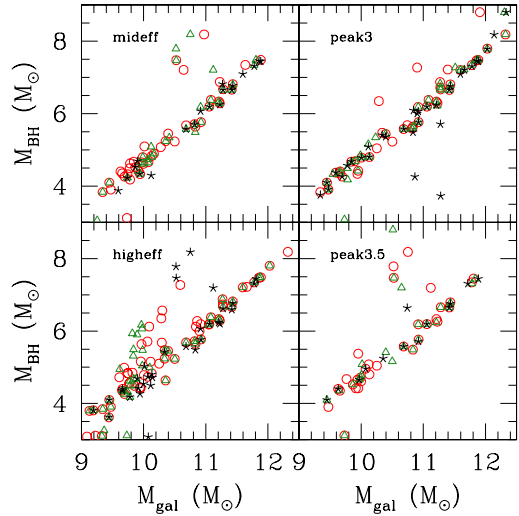}
\caption{ MBH mass against baryonic mass. Circles: no recoil. Triangles: $\hat a=0$. Stars: $\hat a=0.9$. From bottom-left panel, clockwise: model I (higheff), model II (mideff), model III (peak3), model IV (peak3.5).}
\label{fig2}
\end{figure}

An additional prediction that arises from our models is that galaxies which have experienced sustained tidal stripping in the deep potential of the cluster host  MBHs which lie above the expected $M_{BH}-M_{gal}$ correlation. After a merger episode that grew  MBH and galaxy, if the orbital decay of the host brings it in dense regions, tidal stripping will decrease the galaxy mass, while the MBH mass will not be modified. The effect is more pronounced if  MBHs are widespread, as shown in Figure ~\ref{fig2}. If MBHs form at early times, and evolve jointly with their hosts, with accretion bringing the  MBHs exactly on the $M_{BH}-\sigma_*$ relation \citep{DiMatteo2005}, then this population of ``overmassive"  MBHs is unavoidable in galaxy clusters. We strongly stress that our results are robust in a qualitative way, although they should not be assumed quantitatively correct. For instance, the slope of our $M_{BH}-M_{gal}$ correlation is of order $5/3$, because of our assumed rigid scaling between halo properties and  MBH properties.  Regardless of the exact slope, however, the presence of stripped galaxies with overmassive black holes is a robust prediction of models where MBHs form, and grow, early on. We also draw the attention to the possible environmental dependence of the $M_{BH}-M_{gal}$ correlation, but not of the $M_{BH}-\sigma_*$, as the central velocity dispersion is not likely to be affected by the stripping.

There is observational evidence of overmassive  MBHs for a given $M_*$ in Virgo.  \citet{Kormendyetal1997} used ground-based spectroscopic observations of NGC~4486B, the satellite galaxy to NGC~4486 (M87), along with  isotropic dynamical models to find a central dark object of mass $M=6^{+3}_{-2}\times10^8\msun$, though three-integral, axisymmetric models cannot rule out absence of a MBH.   \citet{Magorrian1998} later modeled NGC~4486B with two-integral, axisymmetric dynamical models to find a mass of $M=9.2^{+0.055}_{-0.033}\times10^9\msun$.  The galaxy contains a double nucleus, which may pose problems for isotropic and axisymmetric models, but the presence of a double nucleus, itself, is explained by an eccentric disk of stars, which requires a MBH as in M31 \citep{Tremaine1995}  So while the mass estimate may be uncertain, there is strong evidence for a MBH in NGC~4486B.  Even if the mass estimate is high by a factor of a few, with a stellar mass of $\sim6\times10^9\msun$ \citep{Kormendyetal1997} the MBH is more massive than those in galaxies with comparable bulge mass \citep[see, e.g.,][]{Magorrian1998}. NGC~4486B is almost certainly tidally stripped, though this could be primarily a result of much stronger interactions with M87 \citep{Evstigneevaetal2007} rather than milder stripping from the cluster potential.  Another potential example in Virgo is the S0 galaxy NGC~4342 with a measured black hole mass of $M=3.0^{+1.7}_{-1.0}\times10^8\msun$ and bulge mass of $M=1.2\times10^{10}\msun$ \citep{Cretton1999}, making it another extreme high outlier by a factor of $\sim 30$\citep{Haring2004}.

We note that the population of ``undermassive"  MBHs, signature of the gravitational recoil, described, e.g. in \cite{Volonteri2007, Volonterietal2007} is basically absent here, due to the more simplistic modeling we performed. We have in fact assumed that whenever two galaxies with  MBHs merge, with a mass ratio larger than $1:10$, the  MBHs are placed directly on the $M_{BH}-\sigma_*$. This consequently puts  MBHs preferentially on the $M_{BH}-M_{gal}$ correlation. The cases where the  MBH is below the mass predicted by the $M_{BH}-M_{gal}$ are those in which the galaxy has not had any major merger recently, so the  MBH mass has not grown, while the galaxy mass has \citep[via minor mergers, cfr. the discussion in][]{VHM}. 

\section{Conclusions}
We explored the influence of formation epoch of  MBHs, bias of their hosts, and  MBHs dynamical evolution on the occupation distribution of  MBHs in galaxy clusters. The possible decrease of the occupation fraction at low galaxy mass proposed by \cite{Wehner2006} and \cite{Ferrareseetal2006} is not a surprising result and follows naturally from the evolution of the  MBH population of galaxies in clusters. 
\begin{itemize}
\item When the formation mechanisms of  MBH seeds are taken into consideration, implying formation of  MBHs in massive high redshift halos, the BHOF in cluster galaxies in an increasing function of galaxy mass, in line with observations of the AGN fraction in Virgo. 
\item The exact mass threshold above which BHOF=1 depends on the details of the formation mechanism (host masses and redshift of formation, ``nature") and on the dynamical evolution, including  MBH spin magnitude (``nurture").
\item The repercussions of ``nurture" are magnified in cluster galaxies. If a satellite galaxy looses its  MBH due to a dynamical interaction, it has a negligible chance of capturing a new one following a subsequent galaxy merger, except for the central galaxy in the cluster. 
\item We also predict that if  MBHs co-evolve with galaxies during galaxy mergers, and satellite galaxies experience tidal stripping during their orbital evolution, then a population of galaxies where the  MBH mass lies above the standard $M_{BH}-M_{gal}$ is expected. Two such examples could be NGC~4486B and NGC~4342. 
\end{itemize}

We emphasize that our models do not in any way imply that  MBHs and nuclear clusters are mutually exclusive. We however predict that the occupation fraction decreases with galaxy mass. If, as suggested by Wehner \& Harris (2006), lacking a MBH is a necessary condition for nuclear cluster formation, radio or hard X-ray observations will inform us of the complementary or mutually exclusive essence of  MBHs and nuclear clusters. \cite{Gallo2007} recently observed 32 galaxies in Virgo with the Chandra X-ray Observatory, and found, in agreement with \cite{Decarli2007} that nuclear X-ray activity  increases with the mass of the host galaxy. Intriguingly, at least in one case, VCC1178, a nuclear X-ray source is detected jointly with a central star cluster. \cite{Seth2006} also report that a small AGN might be hosted within the core of the nuclear star cluster in NGC 4206.

\end{document}